\RequirePackage{ifpdf}

\documentclass[aps,twocolumn,nofootinbib,preprintnumbers]{revtex4-1}

\usepackage{aj-definitions1}

\begin{document}

\title{A Mean-Field Description for AdS Black Hole} \author{Suvankar
  Dutta} \email[]{suvankar@iiserb.ac.in} \author{Sachin Shain P.}
\email[]{sachinp@iiserb.ac.in} \affiliation{Department of Physics,
  Indian Institute of Science Education and Research Bhopal \\
Bhopal Bypass, Bhopal-462066}


\begin{abstract}
  In this paper we find an equivalent mean-field description for
  asymptotically $AdS$ black hole in high temperature limit and in
  arbitrary dimensions. We obtain a class of mean-field potential for
  which the description is valid. We explicitly show that there is an
  one to one correspondence between the thermodynamics of a gas of
  interacting particles moving under a mean-field potential and an
  $AdS$ black hole, namely the equation of state, temperature,
  pressure, entropy and enthalpy of both the systems match. In $3+1$
  dimensions, in particular, the mean-field description can be thought
  of as an ensemble of tiny interacting {\it asymptotically flat}
  black holes moving in volume $V$ and at temperature $T$. This
  motivates us to identify these asymptotically flat black holes as
  microstructure of asymptotically $AdS$ black holes in $3+1$
  dimensions.
\end{abstract}


\maketitle

\tableofcontents

\section{Introduction and Summary}\label{intro}


Finding microscopic origin of black hole entropy is an unsolved
problem in theoretical physics. A partial answer has been given in the
context of string theory for a class of asymptotically flat
supersymmetric black holes. However, a complete microscopic
understanding for finite temperature asymptotically flat or $AdS$
black hole is due.

An attempt to construct the microstructure of asymptotically $AdS$
black hole in $3+1$ dimensions has been made in
\cite{Wei:2015iwa}. They introduced the idea of black hole molecules
with some number densities to measure the microscopic degrees of
freedom of the black hole and found that the number density suffers a
sudden change accompanied by a latent heat when the black hole system
undergoes a phase transition. They also found that there is a weak
attractive interaction between two black hole molecules. From these
two phenomena they tried to explore possible microscopic structure of
a charged anti-de Sitter black hole completely from the thermodynamic
viewpoint. However, they could not find what the black hole
microstates actually are. In this letter, we try to speculate about
these black hole molecules.

A seemingly surprising similarity between phase diagrams of an $AdS$
black hole and Van der Waals fluid was first observed in
\cite{myers-catastrophy}. Considering the $AdS$ black hole in a fixed
electric charge ensemble (canonical) and identifying inverse
temperature, electric charge and horizon radius of black hole with
pressure, temperature and volume of liquid-gas system respectively,
\cite{myers-catastrophy} showed that black hole phase diagram is
similar to that of a non-ideal fluid described by the Van der Waals
equation.  However, ad-hoc identifications between the parameters on
both sides were not clear. The story was modified by Kubiznak and Mann
\cite{mann-kubiznak}\footnote{See also \cite{Altamirano:2013uqa}},
who, following \cite{kastor}, considered the negative cosmological
constant to be thermodynamic pressure of the $AdS$ black hole and
volume covered by the event horizon to be thermodynamic volume
conjugate to pressure\footnote{Phase diagram for dyonic black hole has
  been discussed in \cite{suvankar}.}. This allows one to set up an
one to one correspondence between thermodynamic parameters on both
sides. Although remarkable, the analogy between thermodynamics of
charged AdS black holes and that of a Van der Waals fluid was only
qualitative.

The Van der Waals equation, on the other hand, describes an equation
of state of $N$ interacting particles of mass $m$ moving in a volume
$V$ at temperature $T$ under mean field approximation and hence all
the critical exponents of the system take the "mean field" value.
Surprisingly, the critical exponents for $AdS$ black hole evaluated
from the equations of state also take the same "mean field" values. A
natural question arises at this point if there is any effective mean
field description for $AdS$ black hole.

A related question has been attempted to answer in a recent beautiful
paper by Rajagopal {\it et.al.}\cite{Rajagopal:2014ewa}. They
constructed an asymptotically $AdS$ black hole solution whose
thermodynamics matches exactly that of a Van der Waals
fluid\footnote{A generalisation to arbitrary dimensions has been
  constructed in \cite{Delsate:2014zma}.}. However, the black hole
solution they found is "unphysical" because, "the corresponding stress
energy tensor does not obey any of the three standard energy
conditions everywhere outside the horizon, though for certain values
of the parameters it is possible to satisfy the energy conditions in a
region near the horizon.''

\begin{figure}[h]
\centering
\includegraphics[width=8cm,height=5cm]{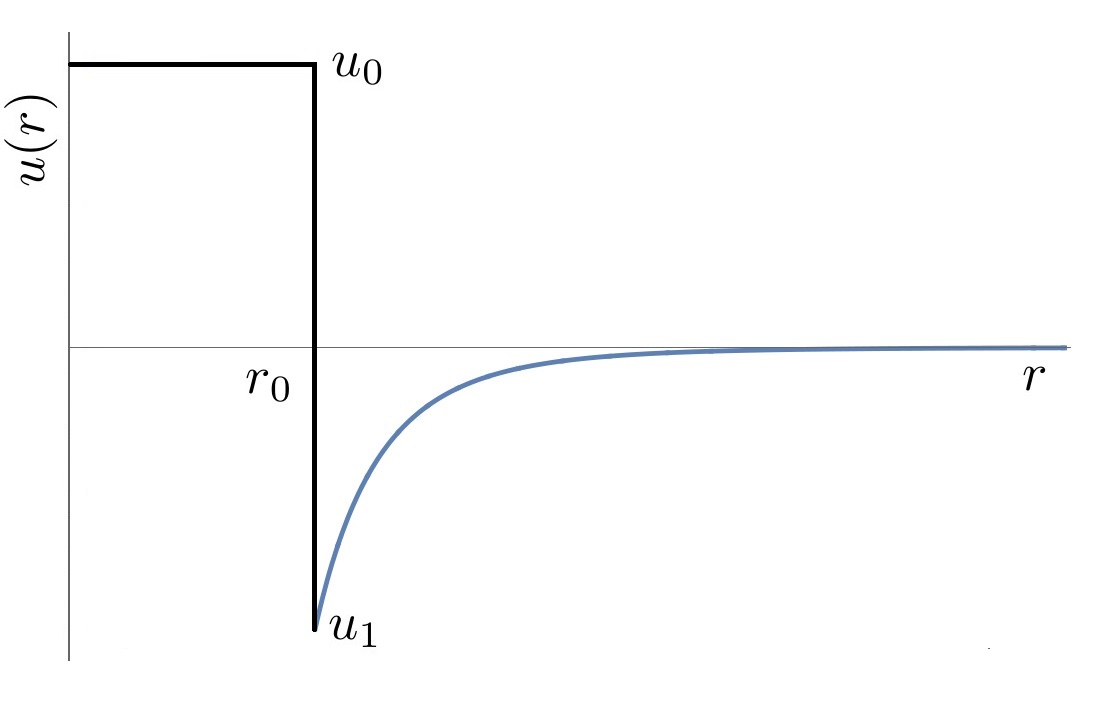}
\caption{A mean-field potential for $AdS$ black hole}
\label{fig:potential}
\end{figure}

We pose a different question. {\it Can one construct an interacting
  system of particles under mean-field approximation such that, not
  only the equation of state, but the complete thermodynamics matches
  with that of an $AdS$ black hole ?}  In this paper we attempt to
answer this question. We find an interacting system of particles of
mass $m^*(T)$ under mean field approximation in $D$ space dimensions
whose equation of motion, entropy, enthalpy and other thermodynamic
quantities match with those of a $D+1$ dimensional $AdS$ black hole in
{\it high temperature limit}. We also find a class of mean-field
potentials for the system of particles. A particular example of such
potential is plotted in figure \ref{fig:potential}.

In $D=3$, in particular, it turns out that the mass of these
interacting objects depends on temperature of the system as
\be
m^*(T) \sim \frac{\hbar c^3}{k G_4 T} + {\cal O}\lb {1\over T^2}\rb.
\ee
This relation is exactly same as the relation between mass and
temperature of an asymptotically flat neutral black hole in $3+1$
dimensions. Therefore, one can think of the particles are essentially
tiny flat black holes moving in a volume $V$. Thus, we can identify
the black hole molecules introduced in \cite{Wei:2015iwa} as tiny
asymptotically flat black holes. This motivates us to find an one to
one correspondence between a gas of interacting flat black holes and
$AdS$ black hole in $3+1$ dimensions and hence the interacting gas
system can be thought of as microstructure of $AdS$ black hole. We
summarise our result in following table.
\begin{center}
\begin{tabular}{|l|l|}
   \hline
\hspace{.6cm} {\bf Mean-field Theory} & \\
{\bf(Interacting black holes} 
& \hspace{.3cm} {\bf $AdS$ Black Hole} \\
\hspace{.4cm} {\bf in 3+1 d flat space)}&\\
\hline
\hspace{.6cm} Temperature & \hspace{.6cm} Temperature\\
\hline
 \hspace{.5cm}  Inverse density &   Inverse charge density\\
\hline
\hspace{1cm} Pressure & Cosmological constant\\
\hline
\hspace{1cm} Entropy & Horizon area\\
&  (entropy of black hole)\\
\hline
\hspace{1cm} Enthalpy & Mass of black hole\\
\hline
\end{tabular}
\end{center}

In other dimensions the equivalence between gas of interacting
particles and $AdS$ black hole holds but the interpretation of
interacting system in terms of an ensemble of flat black holes fails.

The plan of this paper is following. In sec. \ref{sec:CE} we give a
lightning review of classical cluster expansion in arbitrary
dimensions. The mean-field description for $D+1$ dimensional $AdS$
black hole has been discussed in section \ref{sec:4d}. Finally, in
section \ref{sec:concl} we discuss some important features of our
work.

\section{Non-ideal gas in D dimensions and classical cluster
  expansion} \label{sec:CE}

In this section we briefly review the method of cluster expansion for
classical non-ideal gas in $D$ space dimension.  We skip the details
and quote only the main result. A complete discussion on cluster
expansion can be found in \cite{book-huang}.

Let us consider a classical system of $N$ particles of effective mass
$m^*(T)$ moving in a $D$ dimensional volume $V$. The system is in
contact with a heat bath of temperature $T$.  The Hamiltonian is given
by,
\begin{eqnarray}
\mathscr{H} = \sum_{i=1}^{N}\frac{\bf{p}_i^2}{2m^*(T)} 
+ \sum_{i<j}v_{ij} 
\end{eqnarray}
where $\textbf{p}_i$ is total momentum of the $i^{th}$ particle and
$v_{ij} = v(|\textbf{r}_i - \textbf{r}_j|)$ is interaction potential
between $i^{th}$ particle and $j^{th}$ particle. The effective mass
$m^*(T)$ of these particles depends on temperature of the system. The
partition function is given by,
\begin{eqnarray}
 && \mathcal{Q}_N(V,T)= \nonumber\\  
&&\frac{1}{N!h^{DN}}\int d^{DN}p\ 
   \int_V d^{DN}r \exp \bigg[  -\beta\sum_i^N\frac{p_i^2}{2m^*(T)} 
                                - \beta\sum_{i<j}v_{ij} \bigg].\nonumber
\end{eqnarray}
The above integration can be simplified after doing the momentum
integration,
\begin{eqnarray}
\mathcal{Q}_N(V,T) = \frac{1}{N!\lambda^{DN}(T)}\int  
d^{DN}r \text{ exp}\left( - \beta\sum_{i<j}v_{ij} 
\right) \label{c1}
\end{eqnarray}
where $\lambda(T) = \sqrt{2\pi \hbar^2/m^*(T)kT}$ is thermal
wavelength of these particles. The integral in eq.(\ref{c1}) is called
the \textit{configuration integral}.  There is a well known systematic
way of calculating this integral by expanding it in powers of
$(e^{-\beta v_{ij}}$-1). This method is widely known as the
\textbf{cluster expansion} of Ursell and Mayer\cite{cul1}.  The main
application of this method is to calculate the higher order of virial
coefficients for non-ideal gas.

Following \cite{book-huang}, one can compute the grand canonical
partition function in the limit $N\ra \infty$
\begin{eqnarray}
\log \mathcal{L}(z,V,T) = 
\frac{V}{\lambda^D}\sum_{l=1}^{\infty}b_lz^l
\end{eqnarray}
where, $b_l$'s are called {\it cluster integral} and defined as,
\be
b_l(V,T) = \frac{1}{l! \lambda^{Dl-D}V} \label{cluster def}
\text{(sum of all l-cluster)}
\ee
and $z$ is chemical potential. Note that the cluster integrals are
dimensionless. In thermal equilibrium other thermodynamic quantities
like pressure and density of the system can be written in terms of
cluster integrals as,
\begin{eqnarray} \label{eq:Pandv} 
\frac{P}{k T} =
  \frac{1}{\lambda^D}\sum_{l=1}^{\infty} \bbar_l z^l, \quad \frac N V
  = \frac{1}{v} = \frac{1}{\lambda^D}\sum_{l=1}^{\infty} l \bbar_l
  z^l,
\label{c9}
\end{eqnarray}
where
\begin{eqnarray}
\bbar_l(T) \equiv \lim_{V \rightarrow \infty }b_l(V,T).
\end{eqnarray}
From eq.(\ref{eq:Pandv}) one can find that the virial expansion of
equation of state is given by,
\begin{eqnarray} \label{eq:virialexp}
\frac{P v}{k T} = \sum_{l=1}^{\infty} a_l(T) 
\left( \frac{\lambda^D}{v}\right)^{l-1} ,  \label{c10}
\end{eqnarray}
where the virial coefficients $a_l$'s are determined 
in terms of cluster integrals from the following identity,
\begin{eqnarray}
  &&\left (\bbar_1 z + 2 \bbar_2 z^2 + + 3 \bbar_3 z^3 + \cdots \right)
     \bigg[ a_1 + a_2 \left( \sum_{n=1}^{\infty} n \bbar_n z^n \right) 
     \nonumber\\
  && + a_3 \left( \sum_{n=1}^{\infty} n \bbar_n z^n \right)^2 
     + \cdots \bigg] = \bbar_1 z + \bbar_2 z^2 + \bbar_3 z^3 
     + \cdots . \nonumber
\end{eqnarray}
First few of them are given by,
\begin{eqnarray} \label{eos-gas}
a_1 &=& \bbar_1 = 1, \quad a_2 = -\bbar_2, \quad 
a_3 = 4\bbar_2^2 - 2 \bbar_3, \nonumber\\
a_4 &=& -20 \bbar_2^3 + 18 \bbar_2\bbar_3 -
3 \bbar_4, \quad \cdots .
\end{eqnarray}
The entropy of this system is given by,
\begin{eqnarray}
S &=& k T \left(\frac{\partial \log \mathcal{L}}
{\partial T} \right)_{z,V} - N k \log (z) + k 
\log \mathcal{L}\nonumber \\
&=& \frac{k V}{\lambda^D} 
\lB \lb 1- {D T \over \lambda} \frac{d\lambda}{dT}\rb
{\cal B} + T \frac{d{\cal B}}{dT}
\rB
- N k \log (z) \label{cal14}
\end{eqnarray}
where,
\be
{\cal B} = \sum_{l=1}^{\infty} \bbar_l z^l.
\ee
The enthalphy is given by,
\begin{eqnarray}
H &=& U + k T \log \mathcal{L} \nonumber \\
&=& k T^2 \left( \frac{\partial \log \mathcal{L}}
{\partial T} \right)_{z,V} + k T \log \mathcal{L}\nonumber\\
&=&  \frac{k T V}{\lambda^D}\lB \lb 1- {D T \over \lambda} 
\frac{d\lambda}{dT}\rb
{\cal B} + T \frac{d{\cal B}}{dT}
\rB  . 
\label{cal13}
\end{eqnarray}

\section{$D+1$ dimensional electrically charged AdS black hole}
\label{sec:4d}

In this section we derive the equation of state of a $D+1$ dimensional
electrically charged $AdS$ black hole and find an equivalent system of
interacting particles under mean-field approximation whose equation of
state, entropy and enthalpy match with those of black hole.

We consider Reissner-Nordstrom action in $D+1$ dimensions in presence
of a cosmological constant $\Lambda = -\frac{D(D-1)}{2b^2}$
\begin{eqnarray}
I = \frac{c^3}{16 \pi \GD} \int d^{D+1} x \sqrt{-g} \bigg[R  &-&  \frac{\GD
  \eD}{c^2} F^2 \nonumber\\
&+&\frac{D(D-1)}{b^2} \bigg]
\end{eqnarray}
where, $\GD$ and $\eD$ are Newton's constant and permittivity in
$D+1$ dimension. The equations of motion are given by,
\begin{eqnarray}
  R_{\mu \nu} - \frac{g_{\mu \nu}}{2} R -\frac{D(D-1)}{2b^2}g_{\mu
   \nu} & = &
 \frac{2 \GD  \eD}{c^2}(F_{\mu \lambda}F_{\nu}^{\lambda} -
  \frac{g_{\mu  \nu}}{4}F^2),  \label{eq1} \nonumber \\ 
   \nabla_{\mu }F^{\mu \nu}  & =&  0 .\label{eq2} 
\end{eqnarray}
A static, spherically symmetric, electrically charged black hole
solution is given by,
\ben
A & = & \frac1{\eD c \eta}\left(-\frac{q_e}{r^{D-2}} +
  \frac{q_e}{r_+^{D-2}}\right)c dt, \label{asol} \hspace{2.9cm}\\ 
ds^2 & = &-f(r)c^2 dt^2 + \frac{1}{f(r)}dr^2 + r^2d\theta^2 + r^2
       \sin^2\theta d\phi^2 \label{metric}
\end{eqnarray}
where, \be f(r) = \left( 1 + \frac{r^2}{b^2} -\frac{\GD}{c^2} \frac{2
    m}{r^{D-2}} +\frac{\GD}{ \eD c^4 } \frac{q_e^2}{r^{2D-4}} \right)
.\label{fsol} \ee $q_e$ and $m$ are integration constants identified
as electric charge and mass parameters of the black hole.  $r_+$ is
horizon radius. Physical mass and charge are given by,
\be
M= \frac{(D-1)S_{D-1}}{8 \pi}m,\quad Q= \sqrt{2(D-1)(D-2)}
\frac{S_{D-1}}{8 \pi} q_e.
\ee
The asymptotic value of gauge field is identified with the chemical
potential corresponding to charge $q_e$
\begin{eqnarray}
\phie=\frac{1}{\eD \eta} \frac{q_e}{r_+^{D-2}} .\label{phi1}
\end{eqnarray}
Following \cite{mann-kubiznak} we consider the cosmological constant
as thermodynamical pressure of the system
\begin{eqnarray}
  P=-\frac{c^4}{\GD} \frac{\Lambda}{8\pi} =\frac{D(D-1)c^4}{16\pi \GD}
  \frac{1}{b^2}.   
\end{eqnarray}
The hawking temperature of this black hole is given by
\begin{eqnarray}
k T = \frac{(D-2)\hbar c}{4\pi r_+}\left( 1 +  \frac{16 \pi
  \GD}{(D-1)(D-2)c^4} P r_+^2 -
    \frac{\GD \eD}{c^4} \phie^2 \right).\nonumber\\
\label{temp} 
\end{eqnarray}
We consider our system in contact with a reserver at fixed temperature
and chemical potential. Thermodynamic variables associated with our
system are energy $E$, pressure $P$ and charge. Then the first law of
black hole thermodynamics and Helmholtz potential $W$ are given by,
\be
dE = T dS -P dV + \phie dq_e, \quad W = E -T S -\phie q_e
. \label{first law3d} 
\ee
We refer to \cite{myers-catastrophy,suvankar} for detailed
computation of thermodynamic potential and variables. Here we quote
the final results. The free energy is given by, 
\ben
W=\frac{I}{\beta_t} =\frac{S_{D-1}c^4}{16\pi \GD}\bigg[ - \frac{\GD
  \eD}{c^4} \phie^2r_+^D &-&\frac{16 \pi \GD P}{D(D-1)c^4}
r_+^{D} \nonumber \\
&& +r_+^{D-2} \bigg] \label{free} 
\een 
where, $\beta_t$ is periodicity of Euclidean time direction.
$\beta_t = \hbar \beta =\hbar/k T$.  Corresponding thermodynamic
variables can be computed from the free energy,
\begin{eqnarray}
  \qav &=&-\frac{\partial W}{\partial \phie}= \frac{(
           D-2)S_{D-1}}{4\pi} \eD \phi_e r_+^{D-2}, \label{eq:qav}\\
  \vbh  &=&\frac{\partial W}{\partial
            P}=\frac{S_{D-1}}{D}r_+^D,  \label{vol} \\ 
  S_{bh} &=&-\frac{\partial W}{\partial T} \frac{c^3 k }{4\hbar \GD}
             S_{D-1}  r_+^{D-1} = 
        \frac{S_{D-1} k r_+^{D-1}}{4 l_p^{D-1}} 
\end{eqnarray}
where $\lp$ is the Planck's length in $D$ dimensions.  From Helmoltz
free energy we can compute the enthalpy or total energy of the system
as,
\begin{equation}
	H_{bh}=W+TS+\phie q_e = M c^2.
\end{equation} 
Defining $V_o ={D\vbh/S_{D-1}}$, in the limit of large black hole
temperature black hole enthalpy and enthalpy can be written as,
\begin{eqnarray}
\label{enthalpy}
  H_{bh} = {S_{D-1}\over 4 l_p^{D-1}} k T_r  V_o^{{D-1\over D}}, \quad
  S_{bh} = {S_{D-1}\over 4 l_p^{D-1}}  k V_o^{{D-1\over D}} \een 
where
\ben
  \Tpi = \frac{(D-1) }{D}T
\end{eqnarray}
is the reduced temperature.

\subsection{Equation of state}

In this subsection we derive the equation of state of an electrically
charged $AdS$ black hole in $D+1$ dimensions. In thermal equilibrium
the equation of state gives a relation between pressure, temperature
and volume (or density). At high temperature and large volume limit
(classical limit) the equation of state can be written in powers of
inverse temperature and inverse volume.

We first find black hole pressure in terms of volume and temperature
by using eq.(\ref{temp}) and eq.(\ref{vol}),
\begin{eqnarray} \label{eos1}
P=\frac{d}{4 l_p^{D-1}} \frac{k T_r}{V_o^{1/D}} -\frac{(D-1)(D-2)}{16\pi}
\frac{k T_p}{l_p^{D-2}} \lb 1- \frac{\phi^2}{\phi_p^2} \rb
  \frac{1}{V_o^{2/D}} \nonumber \\ 
\end{eqnarray}
where, $\Tp = \hbar c/k l_p$ is Planck's temperature,
$\phip = \frac{q_p}{\eD \eta \lp^{D-2}}$ and 
$\qp=\sqrt{\eD \hbar c l_p^{D-3}}$ is Planck's charge in $D$
dimensions.  

Before we proceed to write the equation of state we notice, from
eq. (\ref{enthalpy}), that both the enthalpy and entropy are
proportional to $V_o^{(D-1)/D}$, in the large volume limit. This is
because of the holographic nature of black hole. Unlike usual
thermodynamic objects entropy and other extensive thermodynamic
quantities of black hole do not depend on volume rather they depend on
the volume of a lower dimensional hypersurface (horizon). Motivated by
this, we define a {\it reduced volume} of the system,
\be \label{redv}
\vred = \lp V_o^{(D-1)/D}.
\ee
We multiply $V_o^{(D-1)/D}$ by a factor of $\lp$ such that the 
reduced volume has the correct dimensions. The entropy and
enthalpy written in term of reduced parameters are given by
\be
\label{bh-ent-enth} S_{bh} = {S_{D-1} k \vred\over 4 l_p^D}, \quad
H_{bh} ={S_{D-1} k \Tpi \vred\over 4 l_p^D}. 
 \ee 
 We further define a quantity reduced volume per unit charge number as
\be
v_r = {\vred\over \qbar} = \lb \frac{\phi_p l_p^{D-1}}{\phi_e}\rb r_+,
\quad \qbar = {q_e\over q_p} = \text{charge number}.
\ee
In the thermodynamic limit, $v_r$ is also very large hence $1/v_r$ can
be considered as an expansion parameter.  Redefining the pressure,
\be\label{pred} \pred = \frac{4\phi_e}{D\phi_p} P, \ee
we can write the equation of states in terms of reduced variables as
\be\label{eos3} \frac{\pred v_r}{ k \Tpi} = 1 - \left[ \frac{(D-1)(D-2)\Tp}
{4\pi D T_r} \lb \frac{\phi_p}{\phi_e} - \frac{\phi_e}{\phi_p}\rb
  l_p^D \right] \lb \frac{1 }{v_r}\rb.  \ee
To match this equation with virial expansion of pressure
(eq.\ref{eq:virialexp}), we introduce black hole thermal wavelength
$\lambda_b(\Tpi)$ and write the equation of state as,
\be\label{eos-final}
\frac{\pred v_r}{ k \Tpi} = 1 - \left[ \frac{(D-1)(D-2)\Tp}
{4\pi D T_r} \lb \frac{\phi_p}{\phi_e} - \frac{\phi_e}{\phi_p}\rb
  {l_p^D \over \lambda_b^D}\right] \lb \frac{\lambda_b^D }{v_r}\rb. 
\ee
Unlike eq.(\ref{eq:virialexp}) this is not an infinite series, the
expansion stops at $l=2$. 

In \cite{mann-kubiznak} the horizon radius, $r_+$, was identified with
the volume per particle, $v$, of a Van der Walls fluid. This
identification was some what ad-hoc. Here we see that $v_r$ which is
reduced volume per unit $charge \ number$ is proportional to $r_+$ and
we identify this inverse charge number density with the inverse
density of Van der Walls fluid.

Our next goal is to set up a dictionary between a thermodynamic system
of $AdS$ black hole and a system of $N$ interacting gas of particles
in volume $V$ and temperature $T$ such that the equation of state,
entropy and enthalpy of these two systems match.

\subsection{The dictionary}

Comparing the equation of state of a charged AdS black hole and that
of a non-ideal gas, we find the following identifications
\ben
\pred = P, \quad v_r = v, \quad \Tpi = T \quad 
\text{and} \quad \lambda_b(\Tpi) = \lambda(T). \ \ \ \ \ \ 
\een
The virial coefficients are given by,
\ben
a_2(T) &=& - \frac{(D-1)(D-2)\Tp}
{4\pi D T_r} \lb \frac{\phi_p}{\phi_e} - \frac{\phi_e}{\phi_p}\rb
  {l_p^D \over \lambda_b^D}, \nonumber\\
  a_l(T)&=&0, \ \ l\ \geq 3.
\een
Therefore, the cluster integrals are given by
\ben\label{b-values}
\bbar_2(T) &=& \frac{(D-1)(D-2)\Tp}
{4\pi D T_r} \lb \frac{\phi_p}{\phi_e} - \frac{\phi_e}{\phi_p}\rb
  {l_p^D \over \lambda_b^D}, \label{4d b2}\\
   \bbar_3(T) &=& 2 \bbar_2^2(T),\quad  
   \bbar_4(T) = \frac{16}{3}\bbar_2^3(T)
  \ \ \cdots.
\een
Equating enthalphy of black hole with that of a non-ideal gas
we get,
\ben \label{lambda-eqn}
 \lb 1- {D T \over \lambda_b(T)} 
\frac{d\lambda_b(T)}{dT}\rb
{\cal B} + T \frac{d{\cal B}}{dT}
 = \frac{S_{D-1} \lambda_b^D(T)}{4l_p^D}.
\een
Finally, equating entropy on both sides we find,
\ben
z=1.
\een
Thus the system of $N$ interacting particles, which is equivalent to
an $AdS$ black hole with constant chemical potential, has constant
fugacity.

Eq.(\ref{lambda-eqn}) along with the values of cluster integrals given
in eq.(\ref{b-values}) can be solved to find $\bbar_2(T)$ and hence
the temperature dependence of the effective mass of an equivalent
system of non-ideal gas of particles. Eq. (\ref{lambda-eqn}) is highly
non-linear and difficult to solve exactly. However, in the limit of
high temperature this equation can be solved and one finds the
temperature dependence of black hole thermal wavelength (or effective
mass) order by order in $1/T$ expansion. At high temperature $\bbar_2$
is small and hence we can neglect the terms proportional to higher
powers of $\bbar_2$. Thus, solving the above differential equation the
thermal wavelength turns out to be
\be
\lambda_b(T) \sim l_p + {\cal O}\lb {1\over T}\rb,
\ee
which implies the effective mass of the particles goes as
\be
m^*(T)= \frac{\hbar^2}{k T} \lb\frac{c^3}{\hbar \GD} \rb^{2\over D-1}.
\ee
In four spacetime dimensions ($D=3$) the relation becomes,
\be
m^*(T) \sim \frac{\hbar c^3}{k G_4 T} + {\cal O}\lb {1\over T^2}\rb.
\ee
This relation is exactly same as the relation between mass and
temperature of an asymptotically flat neutral black hole in four
dimensions. Therefore, we can think of equations of state, entropy,
enthalpy of an interacting gas of asymptotically flat 4 dimensional
black holes in volume $V$ and at temperature $T$ are same as those of
an $AdS$ black hole in constant electric potential ensemble. Thus, we
can identify these asymptotically flat black holes as black hole
molecules introduced in \cite{Wei:2015iwa}. See section
\ref{sec:concl} for a detailed discussion.

\subsection{The mean-field potential}

In this subsection construct the mean field potential in which the gas
of flat black holes are moving. One should note that from
Eqn. (\ref{4d b2}) that,
\begin{eqnarray}
\bbar_2(T) \lambda_b^D = \frac{\alpha}{T}
\end{eqnarray}
where,
$$\alpha  = \frac{(D-1)(D-2)\Tp}
{4\pi } \lb \frac{\phi_p}{\phi_e} - \frac{\phi_e}{\phi_p}\rb
  l_p^D .$$
$\bbar_2$, which is proportional to sum over all 2 clusters
(eq.(\ref{cluster def})), is given by
\begin{eqnarray}
\bbar_2(T)\lambda(T)^D &=& {S_{D-1}\over 2} \int_0^{\infty} r^{D-1}d r \left(
                           e^{-\beta u(r)} -1 \right) =
                           \frac{\alpha}{T}.\ \ \ \ \ \ \ 
\end{eqnarray}
Potentials satisfying the above condition will give
thermodynamic/microscopic description of the $AdS$ black hole at high
temperature. One can choose the following ansatz for the potential,
\begin{eqnarray}
    u(r)=\left\{
                \begin{array}{ll}
                \ \  u_0 & r < r_0\\
              -u_1 \left( \frac{r_0}{r}\right)^n  & r \geq r_0 .\\
                \end{array}
              \right.  \label{potfived}
\end{eqnarray}
To get finite result, $n\geq D$. Assuming $u_0 \beta<<1$ and
$u_1 \beta<<1 $ we find
\begin{eqnarray}
\frac1D \left(u_0-\frac{D u_1}{n-D}\right) r_0^D = -\frac{2\alpha k}{S_{D-1}}.
\end{eqnarray}
The minimum distance of approach $r_0$ can be taken to be twice the size of
a flat black hole.

\section{Discussion} \label{sec:concl}

We find an equivalent system of $N$ interacting particles under
mean-field approximation whose thermodynamics and equation of state
match exactly with those of an electrically charged $AdS$ black hole
in high temperature limit. This qualitative thermodynamic equivalence
between these two systems holds in any arbitrary dimensions.  However,
in four spacetime dimensions, in particular, the interacting system of
particles can be thought of as a classical ensemble of large number of
tiny, interacting asymptotically flat black holes moving in volume
$V$. We consider these black holes as objects/particles moving under a
mean-field potential $u(r)$. Although, these micro black holes
themselves have micro structure, but we do not talk about those
structure, rather we consider them almost point like objects carrying
the degrees of freedom of an $AdS$ black hole. The classical cluster
expansion is valid if $\lambda_b^3/v_r <<1$, that is average inter
black hole distance is much much bigger than the Planck length.

A possible microscopic structure of a $3+1$ dimensional, charged,
anti-de Sitter black hole has been proposed in \cite{Wei:2015iwa} from
the thermodynamic viewpoint. They introduced black hole molecules as
microstructure with a number density and studied how the number
density changes under phase transition between small black hole and
big black hole. They also considered the interaction between these
molecules. Hovere, the exact microstructure was unknown to them. In
this paper we find an one to one correspondence between a gas of
asymptotically flat black holes and $AdS$ black hole and hence we
identify these flat black holes as black hole molecules (as introduced
in \cite{Wei:2015iwa}) of charged $AdS$ black holes in $3+1$
dimensions. The logarithm of phase space volume covered by these tiny
black holes proportional to entropy (horizon area) of $AdS$ black
hole. Note that, our argument is valid at high temperature only where
one can treat the system classically.

\bc {\bf Acknowledgement} \ec 

We thank Akash Jain for initial collaboration. We would like to thank
Nabamita Banerjee for useful discussion. SD acknowledges the Simons
Associateship, ICTP. SD also thanks the hospitality of ICTP, Trieste
where part of this work has been done. Finally, we are grateful to
people of India for their unconditional support towards researches in
basic sciences.

\end{document}